\documentclass[acmsmall,nonacm]{acmart}

\usepackage[english]{babel}
\usepackage{blindtext}

\usepackage{pifont}
\usepackage{float}
\usepackage{subfigure}
\usepackage{tikz}
\usepackage{amsmath,amsfonts}
\usepackage{algorithmic}
\usepackage[ruled,linesnumbered]{algorithm2e}
\usepackage{soul}
\usepackage{multirow}
\usepackage{wrapfig}
\usepackage{dutchcal}
\usepackage{graphicx}
\usepackage{subcaption}
\usepackage{booktabs}
\usepackage{adjustbox}

\usepackage{url}

\newcommand*\circled[1]{\tikz[baseline=(char.base)]{
		\node[shape=circle,draw,inner sep=0.5pt] (char) {#1};}}

\newcommand{\DEL}[1]{\iffalse #1 \fi}

\usepackage{ifthenx}

\usepackage{cleveref}
\crefname{section}{§}{§§}
\Crefname{section}{§}{§§}
\crefformat{section}{§#2#1#3}

\newcommand{\sys}{{\scshape PecSched}\xspace}

\newcommand{\squishlist}{
\begin{list}{$\bullet$}
  { \setlength{\itemsep}{0pt}
     \setlength{\parsep}{0pt}
     \setlength{\topsep}{0pt}
     \setlength{\partopsep}{0pt}
     \setlength{\leftmargin}{0em}
     \setlength{\labelwidth}{0em}
     \setlength{\labelsep}{0.2em} } }

\newcommand{\squishlisttwo}{
\begin{list}{$\bullet$}
  { \setlength{\itemsep}{0pt}
     \setlength{\parsep}{0pt}
    \setlength{\topsep}{0pt}
    \setlength{\partopsep}{0pt}
    \setlength{\leftmargin}{2em}
    \setlength{\labelwidth}{1.5em}
    \setlength{\labelsep}{0.5em} } }

\newcommand{\squishend}{
  \end{list}  }

\begin{document}
% \data{}
%\title{\Sys: A Communication-Efficient Sequence-Parallelism based Serving System for Generative Large Language Models}
\title{\sys: Preemptive and Efficient Cluster Scheduling for LLM Inference}

\author{Zeyu Zhang}
\affiliation{
  \institution{University of Virginia}
  \country{USA}
}

\author{Haiying Shen}
\affiliation{
  \institution{University of Virginia}
  \country{USA}
}
%\titlenote{Produces the permission block, and copyright information}
%\subtitle{Extended Abstract}

% \author{Paper \# XXX, XXX pages}
% \author{Anonymous Author(s)}
% \authornote{Note}
% \affiliation{%
%   \institution{Affiliation}
%   \streetaddress{Address}
%   \city{City} 
%   \state{State} 
%   \postcode{Zipcode}
% }
% \email{email@domain.com}

% \renewcommand{\shortauthors}{X.et al.}

% {\color{red}
% Page limit: must be no longer than 12 pages, including footnotes, figures, and tables. Submissions may include as many additional pages as needed for references and for supplementary material in appendices. 
% }
\newcommand{\sh}[1]{\textcolor{purple}{[HS: #1]}}
\newcommand{\s}[1]{\textcolor{blue}{for HS herself: #1]}}

\begin{abstract}
The scaling of transformer-based Large Language Models (LLMs) has significantly expanded their context lengths, enabling applications where inputs exceed 100K tokens.
Our analysis of a recent Azure LLM inference trace reveals a highly skewed long-tail distribution of input lengths, with approximately 80\% of inputs shorter than 2K tokens. Long inputs constitute only a small fraction.
Existing cluster-level LLM scheduling strategies, including First-In-First-Out (FIFO), reservation-based, and priority-based approaches, primarily target short-input requests with lengths below 2K and fail to address this heterogeneity, leading to inefficiencies such as head-of-line blocking, resource underutilization, and starvation of long-input requests.
We propose \sys, a \underline{P}reemptive and \underline{E}fficient \underline{C}luster \underline{SCHED}uling system for LLM inference. \sys introduces the following key techniques:
1) preemptive scheduling that prioritizes short-input requests for their performance;
2) coordinated prefill-decode colocation and disaggregation, which reduces both the duration and frequency of preemptions;
3) fast Sequence Parallelism (SP) that minimizes the prefill time of long-input requests to further reduce the likelihood and frequency of preemptions.
Evaluations based on Azure LLM inference trace show that, compared to state-of-the-art cluster-level LLM inference schedulers, \sys reduces the 99th percentile queueing delay of short-input requests by up to 92\% and improves their throughput by up to 595\%, without significantly affecting the Job Completion Time (JCT) of long-input requests.
We open-sourced our code.
\end{abstract}

\maketitle

% \color{red}{Used 8 servers and focused on network.}
% \s{ sparsity attention is mainly to reduce computation overhead. In the paper, you need to use more servers, 8? and show the traffic between distant servers takes longer time, so you drop more data for more distant servers, so all messages can arrive at the destinations at the same time --avoiding stragglers. when writing, focus more on network, instead of importance of tokens--these should be studied on KV cache compression, or KV value compression in KV value transmission in the paper Minlan sent to us in meeting}

\section{Introduction}

Transformer-based generative Large Language Models (LLMs) have enabled a wide range of applications. Early use cases such as conversational chat typically involved short-input requests~\cite{sharegpt4}, usually under 2K tokens. As LLMs scale to billions or trillions of parameters~\cite{gpt-2, gpt-3, gpt-4, opt, llama2, llama3.1}, their contextual understanding and ability to process longer, more complex inputs have improved, pushing supported input lengths to 4K-1M tokens~\cite{Bulatov2023ScalingTT, largewordmodel, glm-4, internlm2.5, Gemini}. This expansion has unlocked new applications such as book summarization~\cite{egonmwan2019transformer, zi2022source, zhang2022hegel}, document classification~\cite{adhikari2019docbert, dai2022revisiting}, and code generation~\cite{github-copilot}, where inputs can exceed 100K tokens. 
To accelerate the processing of such long requests, existing work~\cite{loongserve} employs Sequence Parallelism (SP) based on ring attention~\cite{ringattention}.
Our analysis of the recently released Azure LLM inference trace~\cite{azure-llm-trace, azure-llm-trace-analysis} by Microsoft in~\cref{sec:motivation} shows that the input length distribution in today’s LLM inference clusters is a highly skewed long-tail distribution: about 80\% of requests have input lengths below 2K. As the input length increases, the frequency drops.
Compared to input lengths, output lengths are much smaller, implying that input lengths dominate the execution time.
Yet, state-of-the-art cluster-level LLM scheduling strategies~\cite{vllm2023kwon,loongserve,llumnix,exegpt,past-future,fastserve} largely ignore this heterogeneity between short-input and long-input requests.

Cluster-level scheduling in current LLM inference systems generally falls into three categories: First-In-First-Out (FIFO)~\cite{vllm2023kwon, loongserve}, reservation-based~\cite{llumnix}, and priority-based~\cite{exegpt, past-future, fastserve} strategies.
However, these works primarily target short-input requests with input lengths below 2K.
In clusters with both short-input and long-input requests, FIFO~\cite{vllm2023kwon, loongserve} can cause head-of-line blocking, where long inputs delay short ones. Our measurements (in~\cref{sec:motivation}) show that long inputs can increase the 99th percentile queueing delay of short-input requests by up to 10.2×.
Reservation-based strategies~\cite{llumnix} have to pre-allocate sufficient GPUs for long-input requests, limiting the resources available to short inputs and leading to queueing. Since long-input requests are rare, the reserved GPUs often sit idle. Our measurements in~\cref{sec:motivation} show that, compared to FIFO, reservation strategies can increase the 99th percentile delay of short-input requests by up to 1.94×.
Priority-based strategies~\cite{exegpt, past-future, fastserve} can improve the throughput of short-input requests by assigning them higher priority, but this can cause the starvation of long-input requests. Our measurements in~\cref{sec:motivation} show that over 92\% of long-input requests receive no service under such strategies.
FastServe~\cite{fastserve} raises the priority of long-waiting requests but does not specifically target long-input ones. Prioritizing such requests would reintroduce head-of-line blocking.
Critically, due to the variability in input and output lengths, it is difficult to predict them ahead of time, making optimal scheduling inherently challenging.

We propose \sys to address these limitations. \sys is a \underline{P}reemptive and \underline{E}fficient \underline{C}luster \underline{SCHED}uling system for LLM inference.
\sys consists of the following key components.

\textbf{Preemptive scheduling for long-input requests.}
Long inputs increase both the prefill phase (input processing to generate the first token) duration and the iteration time during the decode phase (generating subsequent output tokens).
To ensure the prioritization of short-input requests, unlike prior work~\cite{exegpt,past-future,fastserve} that prioritizes based on output lengths, \sys allows short-input requests to preempt long-input ones, reducing queueing time.

\textbf{Coordinated prefill-decode colocation and disaggregation.}
Since output length is unpredictable at runtime, allowing short-input requests to preempt long-input ones may prolong the suspension time when the decode time of short-input requests is long.
Long decode time of short-input requests preceding a long-input request can also increase the queueing delay of the long-input request.
To address this, we disaggregate the prefill and decode phases of short-input requests. Only the prefill phase is allowed to preempt long-input requests, and long-input requests only need to wait for short-input prefill to complete before execution. Given that prefill is compute-intensive and decode is memory-intensive, we colocate the decode of long-input requests with the prefill of short-input requests. This allows them to run concurrently and avoids suspensions during the decode phase of long input requests.

\textbf{Fast SP for long request prefill.}
Long request prefill phases are more likely to be preempted if their durations are long. To minimize prefill time and reduce the frequency of preemptions, \sys introduces fast SP. Fast SP adopts a hybrid SP strategy. Across nodes, fast SP employs ring attention~\cite{ringattention} to process sequence segments, offering scalable computation with low communication overhead.
However, ring attention has low computational efficiency when the ring is long~\cite{usp-tencent}.
To improve computational efficiency and reduce latency, within a node, fast SP leverages high-bandwidth interconnects (e.g., NVLink) and adopts SP variants that have higher communication volume but superior computational efficiency.

In summary, our work has the following contributions.

\squishlist
\item We analyze a recent Azure LLM inference trace~\cite{azure-llm-trace} and first observe that the distribution of input lengths is highly skewed and imbalanced: the vast majority of requests have short inputs, while long inputs constitute only a small fraction.

\item We propose \sys, a preemptive and efficient cluster scheduling system for LLM inference. \sys employs preemptive scheduling for long-input requests to address the head-of-line blocking problem. It adopts coordinated prefill-decode colocation and disaggregation to reduce both the duration and frequency of preemptions. Additionally, \sys introduces fast SP to minimize long request prefill time, further reducing the likelihood and frequency of preemptions.

%response time for LLM inference and reducing data volume in both KV cache and network communication within the sequence parallelism framework for long prompts.

\item Evaluations based on Azure LLM inference trace demonstrate that, compared to state-of-the-art cluster-level LLM inference schedulers, \sys reduces the 99th percentile queueing delay of short requests by up to 92\% and improves their throughput by up to 595\%, without significantly affecting the Job Completion Time (JCT) of long requests.

\squishend

We open-sourced the code of \sys~\cite{paper-code}.

\vspace{-0.02in}
\section{Background}
\vspace{-0.02in}

\subsection{Existing Cluster-Level LLM Scheduling Strategies}

The cluster-level scheduling strategies adopted by current LLM inference systems can be categorized into three classes: FIFO~\cite{vllm2023kwon,loongserve}, reservation-based~\cite{llumnix}, and priority-based~\cite{exegpt, past-future,fastserve} scheduling.
% Those LLM schedulers are primarily designed for short-sequence requests with request lengths no more than 4K, and do not account for scenarios where short and long sequence requests coexist within the same cluster.

\squishlist
    \item \textbf{FIFO.} FIFO-based LLM schedulers serve requests strictly in the order of their arrival.
    % A key limitation of this approach is that long requests can block the execution of short ones, leading to the head-of-line blocking problem. This increases the queueing delay for short requests and degrades overall serving efficiency.
    \item \textbf{Reservation.} Reservation-based scheduling (also referred to as isolation-based) allocates dedicated resources to different types of requests to eliminate resource interference between them.
    % Under this approach, long requests are assigned significantly more GPU resources than short requests. However, due to the low frequency of long requests, the reserved GPU resources often remain underutilized, leading to inefficient resource usage.
    \item  \textbf{Priority.} Priority-based scheduling assigns higher priorities to short-output requests by analyzing or predicting output length distributions, aiming to improve their throughput.
    % In clusters where short and long requests coexist, prioritizing short requests can indeed enhance their throughput. However, due to the high arrival rate of short requests, the system may continuously serve high-priority short requests, resulting in starvation of long requests, which are unable to obtain GPU resources for execution.
\squishend

% In~\cref{sec:motivation}, we analyze the limitations of each scheduling strategy in detail.
These strategies are primarily designed for requests with input lengths no more than 2K and do not consider the impact of long-input requests. We detail the limitations of these strategies in~\cref{sec:limitation}. 

\subsection{Sequence Parallelism for Long Requests}\label{sec:bkg:sp}

% SP~\cite{ringattention, loongserve} has been introduced in LLM inference for long-input execution to reduce latency and memory pressure by distributing computation and memory usage across multiple GPUs or server nodes. Ring attention~\cite{ringattention} is the most widely adopted method for implementing SP in inference. In this approach, a long input sequence is divided into multiple segments, each processed by a model replica (i.e., a ring attention node). These nodes communicate by passing KV data to collectively compute attention over the full sequence using the online-softmax algorithm~\cite{online-softmax2018milakov}. Within each ring attention node, Tensor Parallelism (TP) can be employed to accelerate the processing of an individual segment~\cite{loongserve}. SP with ring attention enables flexible scaling of model replicas to efficiently serve long-input requests. However, the computation efficiency of ring attention is low, and as the ring length increases, its efficiency degrades, leading to higher latency~\cite{usp-tencent}. Additionally, SP with ring attention only focuses on improving the execution of long-input requests and fails to consider optimizations for workloads where short-input and long-input requests coexist within the same cluster.

SP~\cite{ringattention, loongserve} has been adopted in LLM inference to reduce latency and memory pressure for long-input requests by distributing computation and memory across multiple GPUs or nodes. Ring attention~\cite{ringattention}, the most widely used SP method, partitions a long input into segments processed by model replicas (ring attention nodes), which exchange KV data and compute full-sequence attention via online-softmax~\cite{online-softmax2018milakov}.
Tensor Parallelism (TP) can be applied within each node to accelerate segment processing~\cite{loongserve}. This approach enables flexible scaling for long-input execution. However, its computational efficiency is low and degrades with increasing ring length, resulting in higher latency~\cite{usp-tencent}. Moreover, ring attention is designed solely for long-input execution and overlooks scenarios where short-input and long-input requests coexist.

\section{Motivation}\label{sec:motivation}

In this section, we provide a detailed analysis of the request length distribution in Microsoft’s recently released Azure LLM inference trace~\cite{azure-llm-trace,azure-llm-trace-analysis}, and discuss the limitations of existing LLM scheduling strategies in this context, along with the challenges in addressing them.
For brevity, we refer to short-input and long-input requests as \textbf{short requests and long requests}, respectively, in the rest of the paper when no ambiguity arises.

\subsection{Request Length Distribution}\label{sec:motivation:trace}

\begin{figure}[h]\vspace{-0.16in}
    \centering
    \subfigure[Input distribution.\label{fig:distribution_input}]
    {\includegraphics[width=0.4\columnwidth]{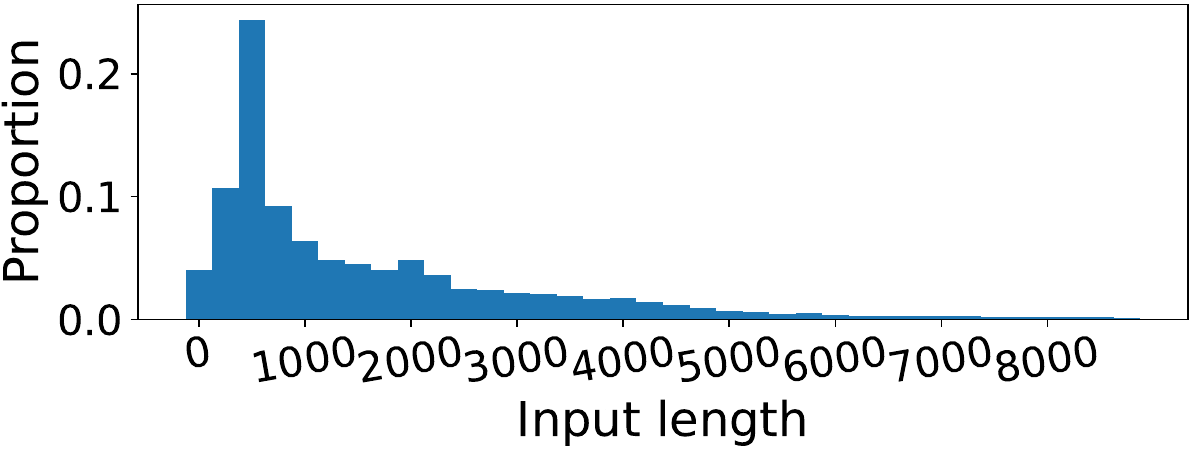}}
    \subfigure[Output distribution.\label{fig:distribution_output}]
    {\includegraphics[width=0.4\columnwidth]{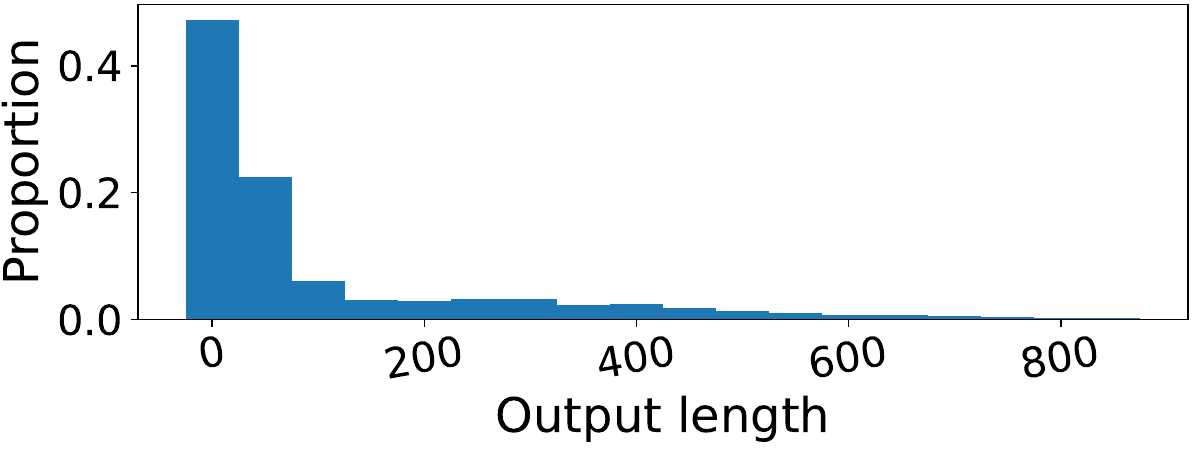}}
    \vspace{-0.22in}
    \caption{Input and output length distributions of Azure LLM inference trace.}
    \label{fig:distribution} \vspace{-0.16in}
\end{figure}

In LLM inference clusters, most requests have short inputs, while long-input requests are rare. Output lengths also vary but remain relatively small. Based on Microsoft’s Azure LLM inference trace~\cite{azure-llm-trace,azure-llm-trace-analysis}, Fig.\ref{fig:distribution} shows a highly skewed long-tail distribution in both input and output lengths. Approximately 80\% of requests have input lengths below 2K, with frequency decreasing as input length grows. This reflects typical usage patterns: short-input tasks (e.g., conversation) dominate, while long-input tasks (e.g., IR\cite{cocktailforir, needlebench}, summarization~\cite{bookcorpus}) are less common. Although output lengths are also long-tailed, they remain under 800 tokens, suggesting that input length has a greater execution-time impact. These observations underscore the need for scheduling mechanisms that account for the heterogeneity between short-input and long-input requests.

\subsection{Limitations in Current Scheduling Strategies}\label{sec:limitation}

The current LLM inference schedulers~\cite{vllm2023kwon,loongserve,llumnix,exegpt, past-future} are primarily designed for short requests with input lengths no more than 2K, and do not account for scenarios where short and long requests coexist within the same cluster.

\begin{figure}[h]\vspace{-0.16in}
    \centering
    \subfigure[Normalized queueing delay of short requests.\label{fig:q_delay_fifo}]
    {\includegraphics[width=0.4\columnwidth]{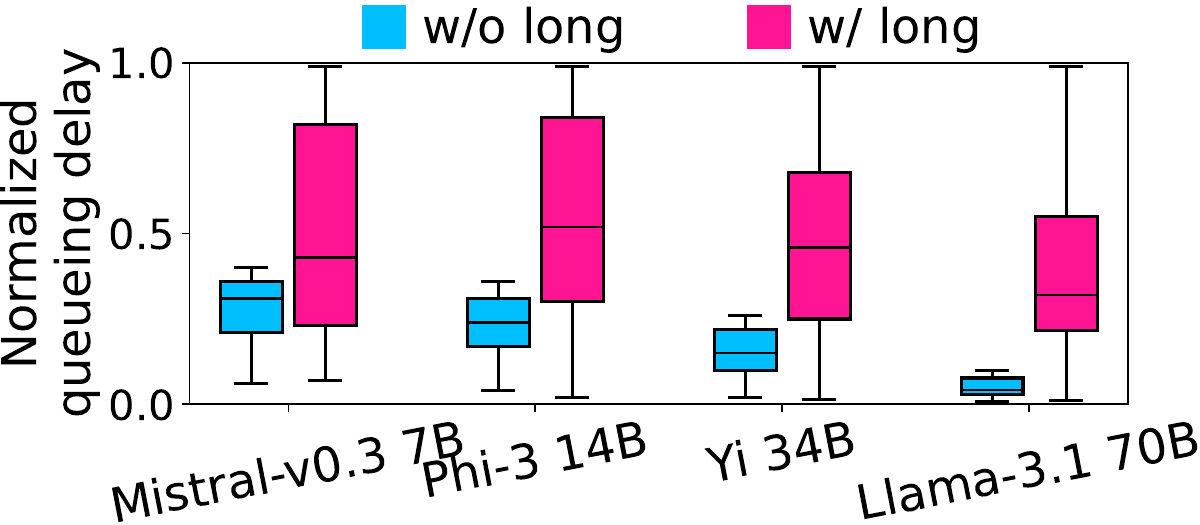}}
    \subfigure[Throughput of short requests.\label{fig:throughput_fifo}]
    {\includegraphics[width=0.4\columnwidth]{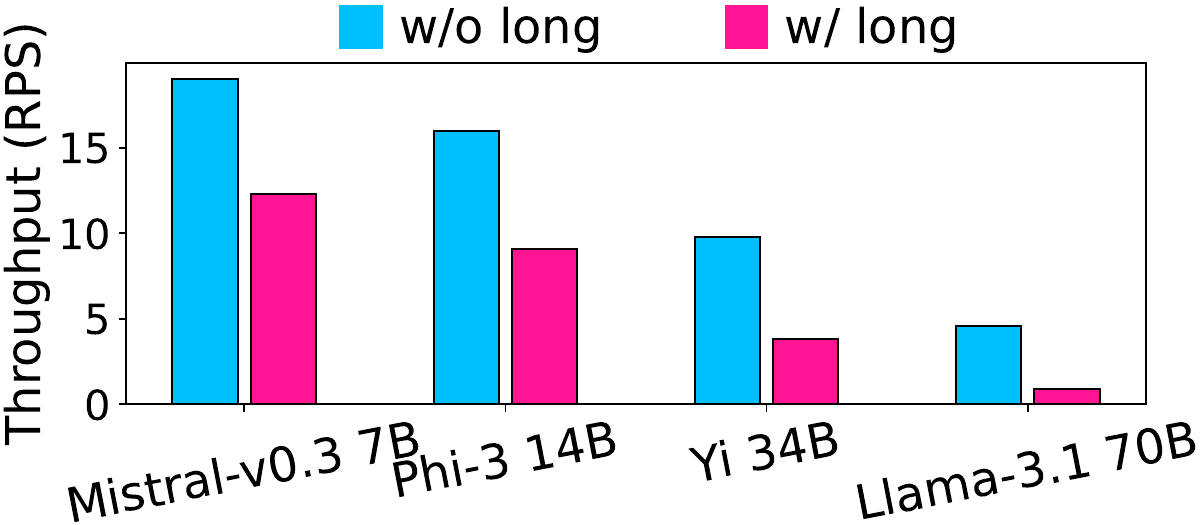}}
    \vspace{-0.22in}
    \caption{The normalized queueing delay and the throughput of short requests with and without long requests when using FIFO.}
    \label{fig:q_delay_throughput_fifo} \vspace{-0.16in}
\end{figure}

\noindent\textbf{FIFO.}
The key limitation of FIFO scheduling~\cite{vllm2023kwon,loongserve} lies in the fact that long requests that require extensive GPU resources and prolonged execution times can block the execution of short requests, leading to the head-of-line blocking problem. This increases the queueing delay for short requests and degrades their throughput.
% The key limitation of FIFO scheduling lies in its handling of long requests, which demand extensive GPU resources and prolonged execution times. As a result, they can hinder the progress of subsequent short requests, causing severe head-of-line blocking.
We evaluate the FIFO-based scheduling strategy using the same setup described in~\cref{sec:eval:setup}.
To study the impact of long requests on head-of-line blocking under a FIFO scheduling strategy, we conduct two experiments using FIFO. In one setting, we remove all long requests from the trace (the definition of long requests is provided in~\cref{sec:eval:setup}), and in the other, we retain all requests. We then evaluate the queueing delay and throughput of short requests under both settings. Fig.~\ref{fig:q_delay_fifo} shows the 1st, 25th, 50th, 75th, and 99th percentile normalized queueing delays of short requests with and without the presence of long requests. We observe that the 99th percentile queueing delay with long requests is 2.5× (Mistral-v0.3 7B), 2.78× (Phi-3 14B), 3.84× (Yi 34B), and 10.2× (Llama-3.1 70B) higher than the delay without long requests. These results indicate that the execution of long requests blocks short requests, significantly increasing their queueing delay. Moreover, as model size increases, the impact becomes more pronounced due to the longer execution time of long requests, resulting in more severe blocking.
Fig.~\ref{fig:throughput_fifo} presents the overall throughput of short requests, measured in Requests Per Second (RPS), under both the presence and absence of long requests. We find that when long requests are present, the throughput of short requests drops to only 0.64× (Mistral-v0.3 7B), 0.56× (Phi-3 14B), 0.39× (Yi 34B), and 0.19× (Llama-3.1 70B) of the throughput observed without long requests. This indicates that the execution of long requests blocks short requests, leading to a reduction in their throughput. Furthermore, the degradation in short request throughput becomes more severe as model size increases, for the same reason discussed earlier.

Since LLM input and output lengths are highly variable (\cref{sec:motivation:trace}) and difficult to predict accurately in advance, it becomes infeasible to proactively partition resources in an optimal way to alleviate head-of-line blocking.
A simple solution is to allow short requests to preempt long requests. However, naive preemption can result in long requests being blocked for a prolonged time by short requests with long execution times. We provide a detailed discussion on how to design more effective preemption mechanisms in~\cref{sec:challenge}.

\noindent\textbf{Reservation.}
In a cluster with many short requests and few long requests, reservation-based scheduling~\cite{llumnix} leads to poor GPU utilization. Substantial resources must be reserved for rare long requests, but these resources often sit idle and cannot be reclaimed by short requests, resulting in significant inefficiency.
This can simultaneously increase the queueing delay and decrease the throughput of short requests.
We evaluate the reservation-based scheduling strategy using the same setup described in~\cref{sec:eval:setup}.

We begin by analyzing the GPU idle rate in the cluster under a reservation-based scheduling strategy. We define the GPU idle rate as:
\vspace{-0in}
\begin{equation}
    \mbox{GPU Idle Rate}=\frac{\sum_i g_i^{I}}{\sum_i (g_i^{E}+g_i^I)},
\vspace{-0in}
\end{equation}
where $g_i^E$ and $g_i^I$ are the total execution time and the total idle time of GPU $i$, respectively.
Table~\ref{tab:gpu_idle_rate} reports the GPU idle rates under the FIFO-based and the reservation-based scheduling strategies for different models.
We observe that under the FIFO strategy, GPUs are rarely idle. In contrast, under the reservation-based strategy, GPU idle rates are significantly higher and increase with model size. This is because long requests require more GPU resources to be reserved than short requests, yet their arrival frequency is relatively low. As a result, a large portion of reserved GPU resources often remains unused, leading to high idle rates. When the model size increases, each long request demands even more GPU resources, necessitating more reservations. This further amplifies resource underutilization and results in higher GPU idle rates.

\begin{table}[h]\vspace{-0.05in}
\centering
\begin{tabular}{ |c|c|c|c|c|  }
 \hline
 GPU idle rate & Mistral-v0.3 7B & Phi-3 14B & Yi 34B & Llama-3.1 70B\\
 \hline
 FIFO & 0.0004 & 0.00009 & 0.0005 & 0.00008\\
 \hline
 Reservation & 0.16 & 0.22 & 0.25 & 0.41\\
 \hline
 \end{tabular}
\caption{GPU idle rate for different models with FIFO and reservation-based strategies.}\label{tab:gpu_idle_rate}
\vspace{-0.25in}
\end{table}

\begin{figure}[h]\vspace{-0.16in}
    \centering
    \subfigure[Normalized queueing delay of short requests.\label{fig:q_delay_resv}]
    {\includegraphics[width=0.4\columnwidth]{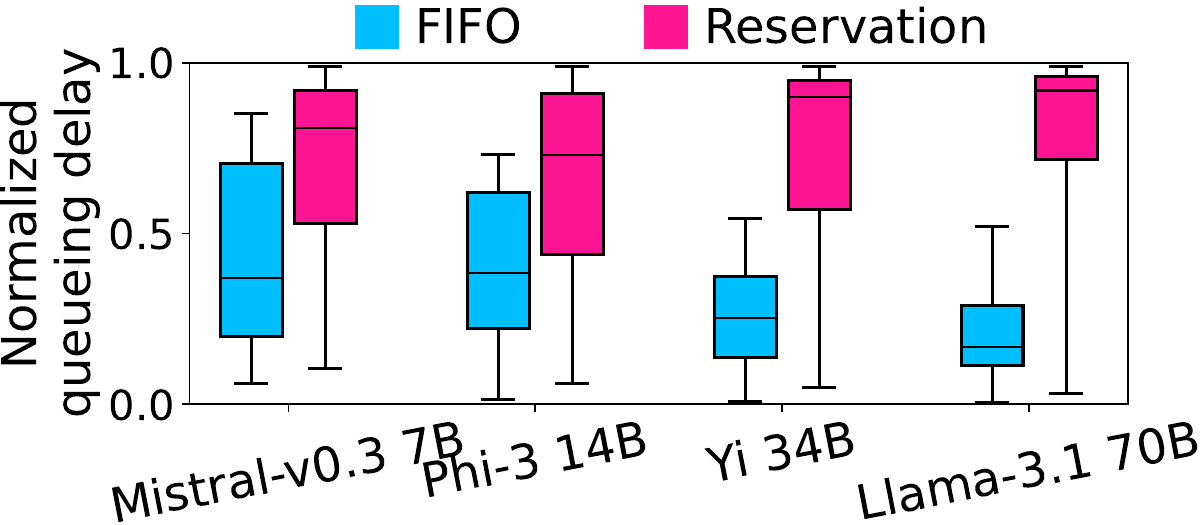}}
    \subfigure[Throughput of short requests.\label{fig:throughput_resv}]
    {\includegraphics[width=0.4\columnwidth]{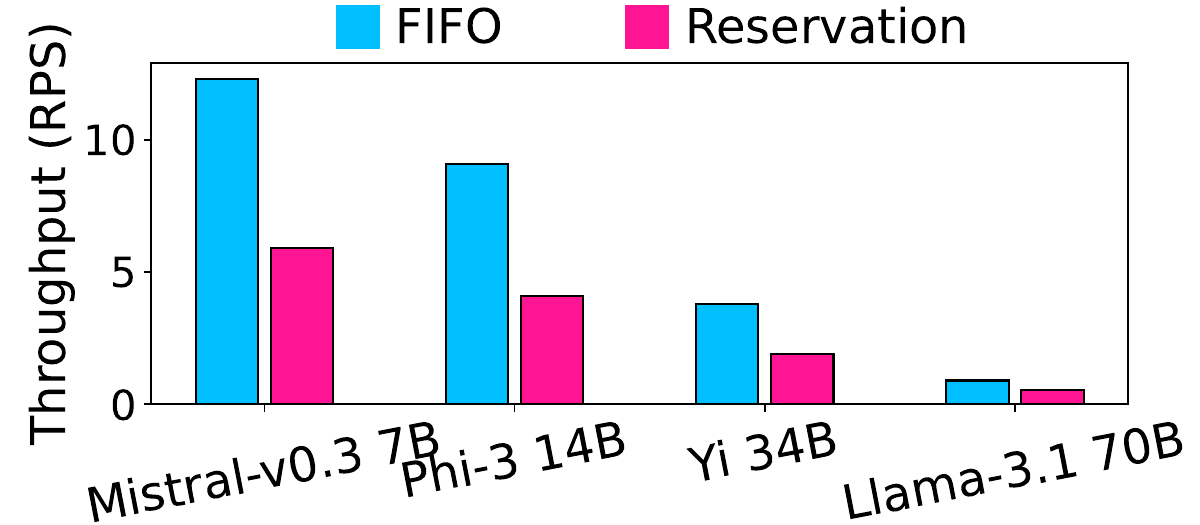}}
    \vspace{-0.22in}
    \caption{The normalized queueing delay and the throughput of short requests when using reservation.}
    \label{fig:q_delay_throughput_resv} \vspace{-0.16in}
\end{figure}

We further analyze the queueing delay and throughput (RPS) of short requests under both FIFO and reservation-based strategies. Fig.~\ref{fig:q_delay_resv} shows the 1st, 25th, 50th, 75th, and 99th percentile normalized queueing delays of short requests. We observe that under the reservation-based strategy, the 99th percentile queueing delay is 1.2× (Mistral-v0.3 7B), 1.35× (Phi-3 14B), 1.8× (Yi 34B), and 1.94× (Llama-3.1 70B) compared to that under the FIFO strategy. This is because the GPU resources reserved for long requests cannot be used to serve short requests, limiting the available resources for them. As the model size increases, the queueing delay of short requests increases further. This is due to the fact that larger models require more GPU resources to be reserved for long requests, which further reduces the GPU resources available for short requests, resulting in longer queueing delays.
Fig.~\ref{fig:throughput_resv} presents the overall throughput of short requests under both FIFO and reservation-based strategies. We observe that under the reservation-based strategy, the throughput of short requests drops to 0.49× (Mistral-v0.3 7B), 0.47× (Phi-3 14B), 0.46× (Yi 34B), and 0.44× (Llama-3.1 70B) of that under the FIFO strategy. The reason is the same as previously explained for Fig.~\ref{fig:q_delay_resv}.

The limitation of reservation-based scheduling can be mitigated by allowing short requests to be dispatched to the GPUs reserved for long requests when those resources are idle. However, this approach still suffers from the same head-of-line blocking issue as FIFO-based scheduling. When an long request is executing, many short requests may still be unable to access GPU resources and are forced to wait.

\noindent\textbf{Priority.}
Priority-based scheduling strategies~\cite{exegpt,past-future,fastserve} assign higher priorities to short-output requests in order to improve their throughput. However, they do not handle long-input requests with a long-tail distribution. Simply assigning higher priorities to short-input requests has the following issue. In clusters where a large number of short-input requests coexist with a small number of long-input requests, the frequent arrival of short-input requests continuously occupies GPU resources. As a result, insufficient resources are left to serve long-input requests, leading to their starvation.
We evaluate the priority-based scheduling strategy using the same setup described in~\cref{sec:eval:setup}.
Table~\ref{tab:starvation_rate} shows the proportion of long requests that are starved under the priority-based strategy.
The vast majority of long requests are starved and never get served. As the model size increases, the proportion of starved long requests also increases. This is because larger models have longer execution times, causing more short requests to accumulate in the queue, which further reduces the chances for long requests to be served.

\begin{table}[h]\vspace{-0.05in}
\centering
\begin{tabular}{ |c|c|c|c|  }
 \hline
 Mistral-v0.3 7B & Phi-3 14B & Yi 34B & Llama-3.1 70B\\
 \hline
 92\% & 97\% & 100\% & 100\% \\
 \hline
 \end{tabular}
\caption{The proportion of long requests that are starved under the priority-based strategy.}\label{tab:starvation_rate}
\vspace{-0.25in}
\end{table}

To avoid starvation of long requests, one potential solution is to promote their priorities~\cite{fastserve}. However, this still introduces head-of-line blocking, as the execution of an long request can delay many short requests queued behind it.

\subsection{Challenges}\label{sec:challenge}

\noindent\textbf{Head-of-line blocking.}
As discussed in~\cref{sec:limitation}, the primary challenge in clusters serving both short and long requests lies in mitigating the head-of-line blocking introduced by the execution of long requests.
Allowing short requests to preempt long requests can mitigate the head-of-line blocking problem. However, naively pausing the execution of long requests through preemption is inefficient for several reasons.
\begin{itemize}
% \squishlist
    \item First, the analysis of the Azure LLM inference trace in~\cref{sec:motivation:trace} reveals that the output lengths of requests are highly imbalanced. While the input length that determines the prefill time is known at the beginning of execution, the output length is unknown, making the decode time unknown at the beginning. Consequently, a preempted long request waits for the slowest short request to complete its decode phase, creating uncertainty in preemption duration and complicating scheduling optimization.
    \item Second, the execution of a request consists of a compute-intensive prefill phase and a memory-intensive decode phase. When a long request is at different phases of execution, how to design preemption strategies that account for its resource usage characteristics becomes a critical question for mitigating the impact of preempting long requests.
\end{itemize}
% \squishend

To address the first issue, we can disaggregate prefill and decode for short requests. Since the decode phase is memory-intensive and short requests have relatively small KV data, we can reserve a small portion of GPU resources exclusively for short request decode. This ensures that only the prefill phase of short requests preempts long requests.
Since prefill time can be estimated from input length at the start of execution, we can leverage this to distribute the prefill workloads of short requests evenly across GPUs during preemption. This approach minimizes the preemption time for long requests by ensuring that short requests complete their prefill phases efficiently in a balanced way during preemption.
Separating the prefill and decode phases of short requests provides an additional advantage. When a long request arrives, it only waits for the ongoing short requests to complete their prefill phases before starting execution, without needing to wait for their decode phases to finish.

% For the second issue, during the prefill phase of an long request, the prefill of short requests can preempt its execution. Since the prefill phases of both types of requests are compute-intensive, the prefill of the long request must be suspended during preemption. During the decode phase of an long request, its KV data volume is significantly larger than that of short requests. To avoid the high communication overhead caused by KV data migration, we retain the KV data on the prefill GPU rather than disaggregating the prefill and decode phases as done for short requests. In this way, since decode is memory-intensive and prefill is compute-intensive, we can colocate the decode of long requests with the prefill of short requests to maximize resource utilization without suspending the decode execution of long requests.

To address the second issue, during the prefill phase of a long request, short request prefills can preempt its execution. As both request types have compute-intensive prefill phases, the long request's prefill is suspended by short request prefills during preemption. In the decode phase, a long request’s KV data volume significantly exceeds that of short requests. To avoid high communication overhead from KV data migration, we keep the KV data of long requests on the prefill GPUs instead of separating the prefill and decode phases as done for short requests. Since decode is memory-intensive and prefill is compute-intensive, we can colocate the decode of long requests with the prefill of short requests, maximizing resource utilization without interrupting the long request’s decode execution.

\noindent\textbf{Reducing preemption of long requests.}
The longer the prefill time of a long request, the more frequently it will be preempted by short request prefills, leading to more preemptions.
Table~\ref{tab:num_preempt} reports the total number of times the prefill phase of all long requests is preempted by short request prefill (i.e., the total number of suspensions). We observe that as model size increases, the number of preemptions experienced by long request prefill increases. This is because larger models result in longer prefill durations for long requests, making them more likely to be preempted by the prefill of short requests.

\begin{table}[h]\vspace{-0.1in}
\centering
\begin{tabular}{ |c|c|c|c|  }
 \hline
 Mistral-v0.3 7B & Phi-3 14B & Yi 34B & Llama-3.1 70B\\
 \hline
 167,394 & 205,947 & 278,504 & 379,305 \\
 \hline
 \end{tabular}
\caption{The total number of preemptions experienced by all long requests.}\label{tab:num_preempt}
\vspace{-0.3in}
\end{table}

To reduce the number of preemptions, it is necessary to shorten the prefill time of long requests, which requires optimizing SP used for their prefill.

As introduced in~\cref{sec:bkg:sp}, state-of-the-art LLM inference systems employ ring-attention-based SP to serve long requests. Each model replica functions as a ring attention node responsible for processing one segment of the input sequence, and TP can be applied within each replica to further accelerate segment processing.
However, relying solely on ring attention and TP is insufficient to minimize prefill time.
The computation efficiency of ring attention is low, and as the ring length increases, its efficiency degrades, leading to higher latency~\cite{usp-tencent}.
SP was originally proposed for training long-context models~\cite{korthikanti2022reducing,deepspeed-ulysses,ringattention}. Megatron~\cite{korthikanti2022reducing} introduced SP to enhance TP performance by reducing memory pressure and increasing the number of tokens that can be processed (thus improving throughput) without increasing communication overhead. Ulysses~\cite{deepspeed-ulysses} further extended SP to reduce the communication overhead in Megatron, lowering latency and improving throughput. However, Ulysses requires that model parameters remain intact (i.e., without TP) during linear projection computations involving model parameters.
Compared to Megatron and Ulysses, ring attention~\cite{ringattention} is more easily scalable, as the number of model replicas can be flexibly adjusted to accommodate sequences of varying lengths, and is more suitable for cross-node communication~\cite{usp-tencent}.
However, despite Megatron and Ulysses having higher communication volume than ring attention, they achieve greater computational efficiency under sufficient bandwidth, resulting in higher throughput and lower latency~\cite{usp-tencent}.
Due to the lack of flexibility compared to ring attention, Megatron and Ulysses have not been adopted in inference systems where requests have varying lengths.

% Based on the characteristics of Megatron, Ulysses, and ring attention, we can adopt a hybrid SP strategy for the prefill phase of long requests in LLM inference to minimize their prefill times. Specifically, ring attention is used to handle cross-node sequence segments due to its better scalability and more efficient communication, while within a node where higher communication bandwidth (e.g., NVLink) is available, Megatron or Ulysses can be employed to improve throughput and reduce latency.
% If the intra-node model replica does not use TP, Ulysses can be applied, whereas Megatron cannot be used without TP~\cite{korthikanti2022reducing}. If the intra-node model replica uses TP, Megatron becomes applicable. Alternatively, Ulysses can still be employed by transmitting model parameters to reconstruct the full parameters required for linear projection computations. In this case, we can estimate both computation and communication costs to select the lower-latency option between Megatron and Ulysses.

To minimize the prefill time of long requests in LLM inference, we can adopt a hybrid SP strategy that leverages the strengths of Megatron, Ulysses, and ring attention. Specifically, ring attention is used across nodes for handling sequence segments due to its scalability and communication efficiency, while within a node, where high-bandwidth interconnects (e.g., NVLink) are available, Megatron or Ulysses is employed to improve throughput and reduce latency.
Megatron requires TP and is applicable only when intra-node model replicas use TP~\cite{korthikanti2022reducing}. In contrast, Ulysses is used without TP and remains viable even when TP is enabled, by transferring model parameters to reconstruct full parameters for linear projections. In such cases, we can estimate both computation and communication costs to choose the lower-latency option between Megatron and Ulysses.

\vspace{-0.04in}
\section{Preliminaries}\label{sec:preliminary}
\vspace{-0.02in}

Before presenting our design, we first introduce some preliminaries on LLMs and SP initially introduced for training. Table~\ref{tab:symbols} lists notations used in the paper.

\begin{table}[h]\vspace{-0in}
\centering
\small
\begin{adjustbox}{max width=\columnwidth}
\begin{tabular}{ |c|l||c|l||c|l||c|l|  }
 \hline
 $d$ & Model dimension size & $N_h$ & The number of heads & $N_l$ & The number of layers & $d_h$ & Head dimension size ($=d/N_h$)\\
 \hline
 $E$ & Token embeddings & $Q$ & Query matrix & $K$ & Key matrix & $V$ & Value matrix\\
 \hline
 $W$& Parameter matrix& $s$& Sequence length & &  & & \\
 \hline
 \end{tabular}
 \end{adjustbox}
 \vspace{-0in}
\caption{Notations used in the paper.}\label{tab:symbols}
\vspace{-0.3in}
\end{table}

\subsection{LLM Foundations}

\begin{figure}[h] \vspace{-0.1in}
    \centering
    \includegraphics[height=2.5cm]{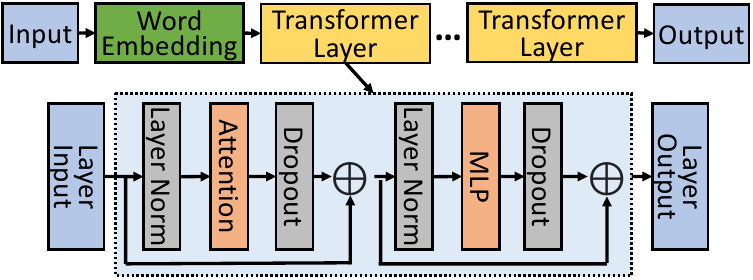}
    \vspace{-0.1in}
    \caption{LLM architecture.}
    \vspace{-0.2in}
    \label{bkg:fig:transformer}
\end{figure}

Fig.~\ref{bkg:fig:transformer} shows the architecture of an LLM, which has multiple transformer layers stacked together. Each transformer layer mainly consists of an attention layer and a Multi-Layer Perception (MLP) layer. An attention layer consists of a QKV generation step, a multi-head self-attention computation, and a post-self-attention linear layer. An MLP block has a linear layer (scaling the model dimension size from $d$ to $4d$), a GeLU layer, and a second linear layer (from $4d$ back to $d$). The output of a transformer layer is the input of the next layer.\looseness=-1

\vspace{0in}

%\subsection{Attention Layer}

\noindent The attention layer in Fig.~\ref{bkg:fig:transformer} includes the components below. 

\noindent \textbf{QKV generation.}
The attention layer takes a sequence embedding $E$ as an input. QKV generation conducts the following operations to generate $Q$, $K$, and $V$ of the sequence for head $h$:
\vspace{-0in}
\begin{equation}\label{equ:qkv}
    Q^h=EW_Q^h \mbox{, ~} K^h=EW_K^h \mbox{, ~} V^h=EW_V^h, \vspace{-0in}
\end{equation} where $W_Q^h$, $W_K^h$, and $W_V^h$ are the parameters for QKV generation in the attention head $h$.

\noindent \textbf{Self-attention.}
The self-attention layer takes $Q^h$, $K^h$, and $V^h$ as inputs, and outputs
\vspace{-0in}
\begin{equation}\label{eq:attention}
    % O^h=P^hV^h \mbox{ for } P^h=\mbox{Softmax}(\frac{Q^h(K^h)^T}{\sqrt{d_h}})\vspace{-0.05in}
    O^h=\mbox{Softmax}(\frac{Q^h(K^h)^T}{\sqrt{d_h}})V^h=P^hV^h, \vspace{-0in}
\end{equation} where $P_h$ is the attention probability for head $h$. The softmax function operates row-wise on the input matrix $[a_{i,j}]$ as follows: \vspace{-0.1in}

\vspace{-0in}
\begin{equation}\label{eq:softmax}
    % \begin{split}
        \DEL{[b_{i,j}]=\mbox{Softmax}([a_{i,j}])\mbox{, for }
        b_{i,j}=}
        \frac{exp(a_{i,j})}{\sum_{k=1}^{t_i}exp(a_{i,k})}, \vspace{-0in}
    % \end{split}
\end{equation}where $t_i$ is the index of the token on row $i$.

\noindent \textbf{Post-self-attention linear.}
The post-self-attention linear layer takes $O^h$ from all heads as the input, and it outputs
\vspace{-0.1in}
\begin{equation}\label{eq:finaloutput}
    % \begin{split}
        O_L = [O^1,O^2,...,O^{N_h}]W_L=OW_L, \vspace{-0in}
    % \end{split}
\end{equation} where $O$ is a concatenation of $O^h$, and $W_L$ is the parameter of the post-self-attention linear layer.

\subsection{Sequence Parallelism of Megatron and Ulysses}\label{sec:pre:sp}

\begin{figure}[h]\vspace{-0.16in}
    \centering
    \subfigure[Megatron SP.\label{fig:megatron}]
    {\includegraphics[width=0.495\columnwidth,height=3cm]{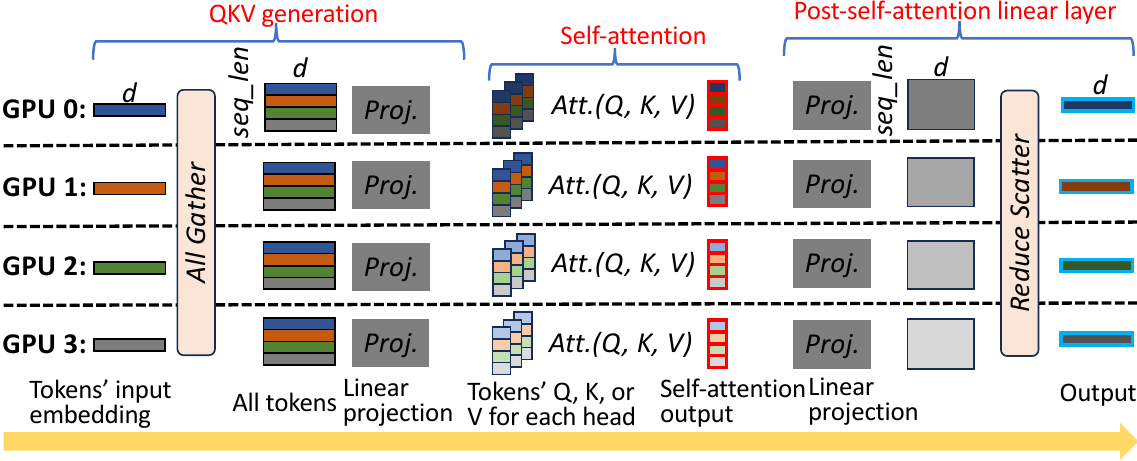}}
    \subfigure[Ulysses SP.\label{fig:ulysses}]
    {\includegraphics[width=0.495\columnwidth,height=3cm]{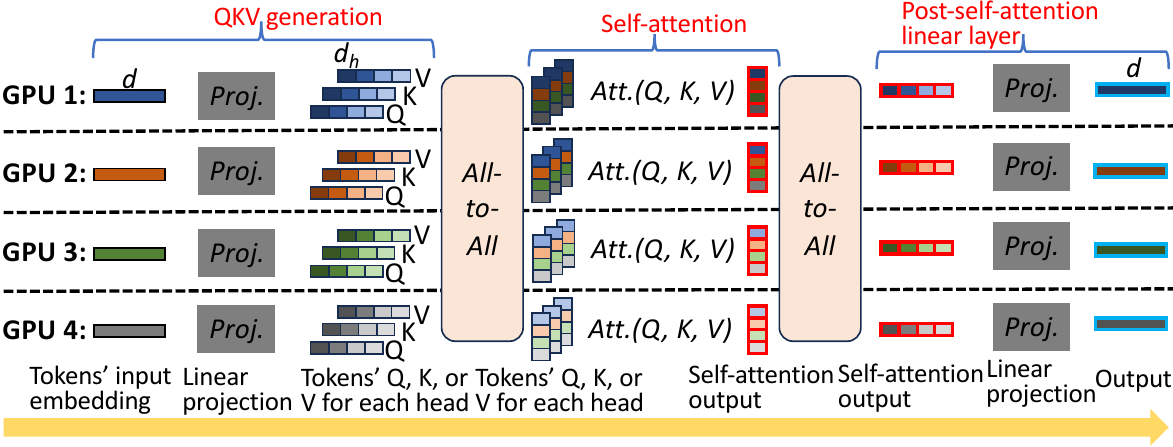}}
    \vspace{-0.22in}
    \caption{Megatron and Ulysses SP architectures.}
    \label{fig:sp_basics} \vspace{-0.16in}
\end{figure}

We introduce the SP architectures used by Megatron and Ulysses.
Fig.~\ref{fig:megatron} illustrates Megatron SP. Each GPU has a different sequence segment's token embeddings (aka, hidden states) as the input to attention. Each GPU executes Eq.~\eqref{equ:qkv} for the first projection to calculate the QKV data of its segment for all heads. Then, the heads are evenly split across GPUs for attention computation. For this purpose, the first all-to-all (A2A) communication sends different head partitions' QKV data from a GPU to their assigned GPUs. After the first A2A, each GPU has the QKV of the entire sequence for part of the heads, and then executes self-attention in Eq.~\eqref{eq:attention}.
Next, it first gathers the head dimension and splits the sequence dimension of the self-attention output through the second A2A. Then, it conducts the second projection to get the final output via a linear matrix transformation as in Eq.~\eqref{eq:finaloutput}.

Fig.~\ref{fig:ulysses} illustrates Ulysses SP. At the beginning, four GPU devices have their sequence segments. Before the first linear projection, due to the use of TP, each GPU must collect all segments through all-gather. Then, each GPU executes the first projection (Eq.~\eqref{equ:qkv}) to obtain the QKV data of the entire sequence of its assigned heads since each GPU is responsible for certain heads (i.e., one head in this example) in TP. Then, each GPU conducts the attention computation on its QKV data (Eq.~\eqref{eq:attention}) and outputs the self-attention result for that head partition.
The second linear projection in the post-self-attention linear layer conducts a linear matrix transformation between the self-attention output $O^h$ and part $i$ of the parameters of the linear layer $W_L^i$, generating $O^hW_L^i$, a $L_{in}$-by-$d$ matrix.
Next, using reduce-scatter, $O^hW_L^i$ from all GPUs are added together to obtain the $O_L$ in Eq.~\eqref{eq:finaloutput} and then split in sequence dimension into each GPU.

\section{Design}\label{sec:design}

Our system consists of the following main components. First, we enable preemption of long request prefill by short request prefill to mitigate head-of-line blocking. Second, we introduce coordinated prefill-decode colocation and disaggregation to disaggregate short request prefill and decode, and colocate long request decode with short request prefill to reduce the impact of preemptions on long requests. Third, we introduce fast SP for long request prefill by employing a hybrid SP strategy to accelerate execution and reduce the frequency of preemptions.

\begin{figure}[h] \vspace{-0in}
    \centering
    \includegraphics[height=3cm]{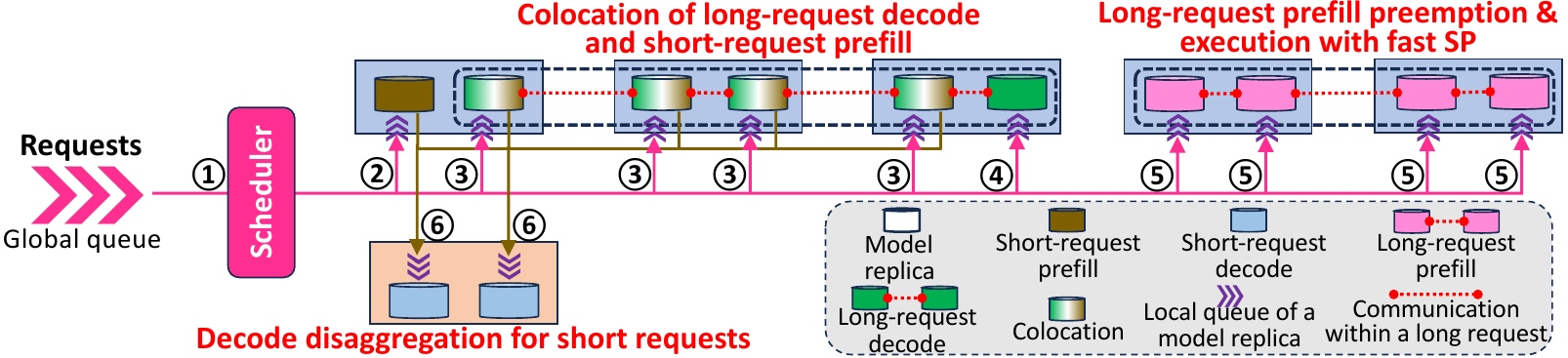}
    \vspace{-0.1in}
    \caption{Design overview.}
    \vspace{-0.16in}
    \label{fig:overview}
\end{figure}

Fig.~\ref{fig:overview} provides an overview of our system architecture.
Requests first arrive at the cluster-wide global queue and are dispatched by the scheduler (step \circled{1}). For short requests that can be handled by a single model replica, the scheduler first attempts to place them in the local queues of idle replicas not occupied by long request prefill and decode (step \circled{2}). If no such replicas are available, it then considers colocating them with long request decode on GPUs with available compute capacity (steps \circled{3} and \circled{4}). If neither is feasible, the scheduler preempts long request prefill to make room for short request prefill execution (step \circled{5}). Once the short request prefill completes, its KV data is migrated to a dedicated decode-only node for decode (step \circled{6}). Long requests, on the other hand, are scheduled across a sufficient number of model replicas, prioritizing those within the same node to minimize communication overhead.
The local queue length of a model replica is defined by the number of tokens in the queue~\cite{splitwise}. When a request can be served by multiple valid combinations of model replicas, the combination with the smallest total local queue length is selected.

\vspace{0in}

\subsection{Preemption of Long-Request Prefill}

While a long request prefill is executing, newly arrived short requests may accumulate. When these short requests need to preempt the ongoing prefill, the system pauses the long request prefill and saves all necessary data for future resumption. Since an LLM is composed of a stack of identical transformer layers, we categorize the data to be saved into two parts: data from completed layers and data from the currently executing layer.

For completed layers, only the generated KV data needs to be preserved for the future decode phase. For the layer currently being executed, we first check whether the KV data for that layer has been generated. If so, the KV data must be retained. We then identify the precise pause point in the prefill execution, which always falls between two kernel operations. A kernel operation may involve computations such as linear projections or self-attention. In such cases, we only need to store the intermediate data passed between these two operations, which is the output of the previous kernel or the input to the next. This intermediate data is usually the token embedding.

Since only one layer’s intermediate data needs to be stored, the memory footprint of this data is small, usually less than 5\% of the total size of all KV data. To resume execution, the system simply continues from the pause point using the stored intermediate data.

\subsection{Coordinated Prefill-Decode Colocation and Disaggregation}

\noindent\textbf{Prefill-decode disaggregation for short requests.}
Since short requests typically generate a small amount of KV data and their decode phase is computationally lightweight, we reserve only a small subset of model replicas to handle short request decode. The KV data produced during the prefill phase is transmitted to these decode-only nodes for decode. To further reduce KV transmission latency, we overlap prefill computation with KV transmission. As soon as the KV data for a transformer layer is generated, it is immediately sent to the decode replica, overlapping with the computation of the next layer’s prefill.

Because short request prefill and decode are disaggregated, we only need to preempt long request prefill with short request prefill. Since the input length of a short request is known at the time of arrival, we can estimate its prefill latency. To ensure load balance across model replicas during preemption, we construct short request batches such that the total number of tokens per batch is balanced across replicas.

\noindent\textbf{Colocation of long request decode with short request prefill.}
When a long request enters the decode phase, it typically processes only one input token at a time, resulting in underutilization of GPU compute resources. To improve utilization without suspending long request decode, we colocate short request prefill computations with long request decode. Since the KV data generated during short request prefill is proactively transmitted to the decode node as soon as it is produced, rather than waiting for the entire prefill phase to complete, it does not consume additional GPU memory. This makes it feasible to colocate long request decode with short request prefill on the same GPU.

\begin{figure}[h] \vspace{-0.1in}
    \centering
    \includegraphics[width=0.99\columnwidth]{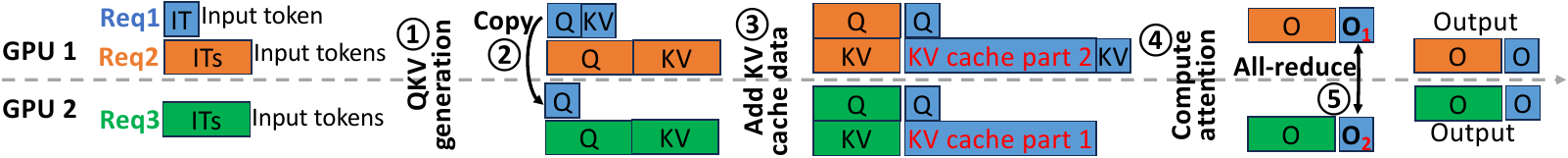}
    \vspace{-0.1in}
    \caption{An example of colocation of long decode with short prefill.}
    \vspace{-0.16in}
    \label{fig:colocation}
\end{figure}

Fig.~\ref{fig:colocation} illustrates an example of such colocation. The setup involves two GPUs. Req1 is a long request undergoing decode, containing only a single input token. Req2 and Req3 are short requests in their prefill phase, each with multiple input tokens. Req1 and Req2 are colocated on GPU 1, while Req3 is assigned to GPU 2. All input tokens are first processed through QKV generation as indicated in Eq.~\eqref{equ:qkv} to obtain their respective $Q$, $K$, and $V$ data (step \circled{1}). Req1’s $Q$ is then copied from GPU 1 to GPU 2 for self-attention computation with the cached KV data (step \circled{2}). Since Req1 only has one token, the communication overhead is small. Meanwhile, the new KV data generated for Req1 on GPU 1 is concatenated with its existing KV cache (step \circled{3}). Each GPU then performs self-attention as indicated in Eq.~\eqref{eq:attention} using the local $Q$ and KV data (step \circled{4}), producing the output $O$ for each request. Req1's self-attention outputs from GPU 1 and GPU 2 ($O_1$ and $O_2$) are merged via an all-reduce operation to form the final output $O$ (step \circled{5}).

\vspace{0in}

To avoid introducing additional latency to long request decode, the scheduler balances the total number of input tokens across GPUs and constrains the token count per GPU to a threshold that ensures no degradation in decode performance.

\subsection{Fast Sequence Parallelism for Long-Request Prefill}

To further reduce the prefill time of long requests and decrease the frequency of preemption by short request prefill, we aim to improve the performance of SP during the prefill phase. Specifically, we leverage the high intra-node bandwidth to enhance inter-GPU collaboration, thereby improving token processing efficiency and reducing overall processing latency. Across nodes, we adopt ring-attention-based KV transmission for SP computation to minimize inter-node communication overhead while maintaining flexibility. Within each node, we employ a hybrid strategy combining Megatron SP and Ulysses SP, which are introduced in~\cref{sec:pre:sp}, and dynamically select the fastest approach based on input lengths to minimize the prefill time.

\begin{figure}[h] \vspace{-0in}
    \centering
    \includegraphics[width=0.99\columnwidth]{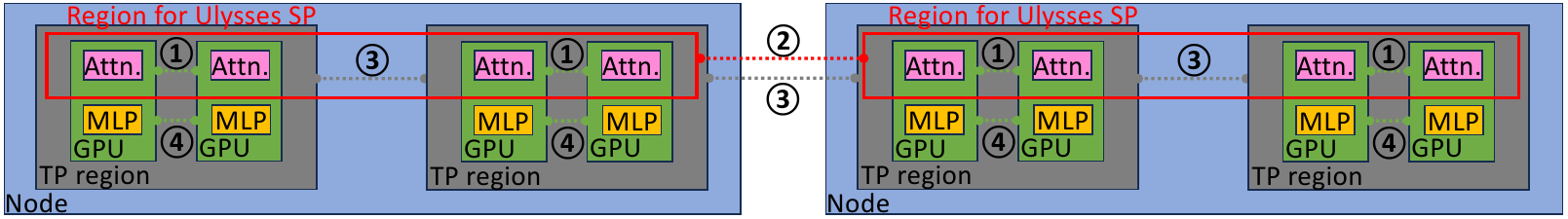}
    \vspace{-0.1in}
    \caption{An example of fast SP for prefill with a hybrid SP strategy.}
    \vspace{-0.16in}
    \label{fig:fast_sp}
\end{figure}

Fig.~\ref{fig:fast_sp} illustrates an example of fast SP employing a hybrid SP strategy. Fast SP consists of two main stages: the attention stage and the MLP stage. During the attention stage, the system selects either Megatron SP or Ulysses SP within each node based on the actual sequence length to achieve the fastest prefill time. When a model replica uses TP (referred to as a TP region), Ulysses SP requires parameter transmission (step \circled{1}) to ensure that each GPU holds the complete parameters needed for computation. This transmission overhead is considered in the strategy selection process. Across nodes, computation is performed using ring attention. If Ulysses SP is used, the KV transmission for ring attention occurs across nodes (step \circled{2}); if Megatron SP is used, the KV transmission for ring attention occurs between TP regions (step \circled{3}).
In the MLP stage, if TP is used without parameter transmission for reconstructing full parameters, the Megatron SP is adopted. Specifically, token embeddings are all-gathered before the MLP, and the MLP outputs are reduced-scattered afterward~\cite{korthikanti2022reducing}. Alternatively, if parameters are transmitted (step \circled{4}) such that each GPU holds the complete parameters, MLP computation can be performed directly on each sequence segment without any token embedding communication~\cite{deepspeed-ulysses}.

\vspace{0in}

We describe how to select the SP strategy within a node. Let the TP size be denoted by $T$, the number of GPUs in a node be $G$, and the sequence segment length processed by each GPU be $s$. Assume that Group Query Attention (GQA)~\cite{group-query-attention} uses $N_h$ query heads and $N_h^{KV}$ key-value heads.
In the attention stage, if Megatron SP is selected, the total communication volume in a node is $2sd(T-1)G$ for all-gather and reduce-scatter, and the total computation volume on a GPU is $2sd(N_h+N_h^{KV})d_h/T + 4(sT)^2d/T + 2sd^2$ for QKV generation, self-attention, and post-self-attention linear layer; if Ulysses SP is selected, the total communication volume in a node is $2s(N_h+N_h^{KV})d_h(G-1) + (d(N_h+N_h^{KV})d_h+d^2)G(T-1)/T$ for two A2A communications and parameter communications, and the total computation volume on a GPU is $2sd(N_h+N_h^{KV})d_h + 4(sG)^2d/G + 2sd^2$ for QKV generation, self-attention, and post-self-attention linear layer.
In the MLP stage, selecting Megatron SP results in communication volume $2sd(T-1)G$ in a node for all-gather and reduce-scatter, and computation volume $16sd^2$ on a GPU for linear projections; selecting Ulysses SP results in communication volume $8d^2(T-1)G/T$ in a node for parameter communications, and computation volume $16sd^2$ on a GPU for linear projections.
There are four possible combinations of SP strategies across the two stages. For each combination, we compute and estimate the total communication time and computation time, and select the one with the lowest overall latency.

\vspace{-0in}
\section{Performance Evaluation}\label{evaluation}
\vspace{-0in}

\subsection{Implementation}\vspace{-0in}

We built our system on vLLM~\cite{vllm2023kwon}. We extended the model code with our custom DistributedAttention class. %SP and \methodone. 
We used the Triton~\cite{triton} version of FlashAttention-2~\cite{flashattention-2} as the core attention but modified it to support the hybrid SP strategy. We modified the LLMEngine and DistributedGPUExecutor in vLLM to enable SP and launch Ray~\cite{ray} workers for SP partitions. The communication backend is NVIDIA Collective Communications Library (NCCL)~\cite{nccl}.
Each worker handling a sequence segment on a GPU runs as a Ray worker and communicates with each other via PyTorch's \textit{distributed} package with the NCCL backend.
The implementation of short request prefill-decode disaggregation was built upon the disaggregation framework in vLLM, but we further extended it by enabling overlap between KV transmission and prefill computation.

\vspace{-0in}
\subsection{Experiment Setup}\label{sec:eval:setup}

In the evaluation, unless otherwise specified, we utilized the following settings.

\noindent\textbf{Testbed.}
We employed four AWS p4de.24xlarge instances \cite{awsp4} located in four nodes. Each instance is equipped with 8 NVIDIA A100 GPUs (each with 80 GiB memory), 96 vCPUs, and 1152 GiB host memory, connected with a 400 Gbps network.

\noindent\textbf{Models.}
Table~\ref{tab:models} lists the state-of-the-art models we used with their TP and Pipeline Parallelism (PP) size, following the setting in~\cite{Agrawal2023SARATHIEL, distserve}.

\begin{table}[h]\vspace{-0.1in}
\centering
\small
\begin{adjustbox}{max width=\columnwidth}
\begin{tabular}{|c|c||c|c|}
\hline
Mistral-v0.3 7B~\cite{mistral-v0.3} & no TP and PP & Phi-3 14B~\cite{phi-3} & no TP and PP \\
\hline
Yi 34B~\cite{yi-model} & TP=4, no PP & Llama-3.1 70B~\cite{llama3.1} & TP=4, no PP \\
\hline
\end{tabular}
\end{adjustbox}
\vspace{-0in}
\caption{Model size and TP/PP size.}
\vspace{-0.3in}
\label{tab:models}
\end{table}

\noindent\textbf{Trace.}
We used the Azure LLM inference trace~\cite{azure-llm-trace} published by Microsoft Azure to generate requests. We adhered to the request arrival times recorded in the trace.
We mimicked the input and output length distributions observed in the trace.
% While datasets such as IR~\cite{cocktailforir,needlebench} and book summarization~\cite{bookcorpus} contain inputs as long as 100K–500K tokens, the maximum input length in the real trace is only around 9K. To better reflect realistic usage in emerging applications, we treated requests with input lengths above the 95th percentile in the trace as long-input requests and replaced their input lengths with values randomly sampled from 100K to 500K. Other requests were short requests. This is motivated by the input length trend observed in the real trace (\cref{sec:motivation:trace}): most requests have short inputs, while a small fraction involve long inputs.
Since the maximum input length in the trace is only around 9K, while common long-input datasets such as IR~\cite{cocktailforir,needlebench} and book summarization~\cite{bookcorpus} contain inputs as long as 100K–500K tokens, we adjusted the trace to better reflect such long-input workloads. Given that the proportion of requests decreases with increasing input length, resulting in long-input requests constituting only a small fraction (\cref{sec:motivation:trace}), we classified requests with input lengths above the 95th percentile in the trace as long-input requests and replaced their input lengths with values randomly sampled from the range of 100K to 500K. All remaining requests were treated as short-input requests.
As for output lengths, since the trace and the common long-input datasets such as IR and book summarization exhibit similar output lengths ranging from tens to a few hundred tokens, we directly mimic the output length distribution in the trace without modification.

\noindent\textbf{Comparison methods.}
We adopted the following three comparison methods.
\begin{itemize}
% \squishlist
    \item \textbf{FIFO.} We used vLLM~\cite{vllm2023kwon} as the comparison method that adopts a FIFO-based scheduling strategy.
    \item \textbf{Reservation.} We used Llumnix~\cite{llumnix} as the comparison method that adopts a reservation-based strategy. Specifically, it pre-allocates GPU resources capable of handling requests with input lengths of 500K tokens, dedicating them to serve long requests with input lengths between 100K and 500K. The remaining GPU resources are reserved exclusively for serving all other short requests.
    \item \textbf{Priority.} We used Past-Future~\cite{past-future} as the comparison method that adopts a priority-based scheduling strategy. Requests with input lengths between 100K and 500K are assigned low priority, while all other short requests are given high priority.
\end{itemize}
% \squishend

For long requests, if a single model replica is insufficient to serve a request, SP with ring attention introduced in~\cref{sec:bkg:sp} is employed to partition the sequence and distribute the computation across multiple model replicas.

\noindent\textbf{Scheduling.}
A cluster has a global queue, and each model replica in the cluster has a local queue.
Each comparison method schedules all requests in the global queue based on its scheduling strategy.
A selected request from the global queue is sent to a model replica that has the shortest local queue length~\cite{splitwise}, defined by the number of tokens. Local queues adopt FIFO.
If a request requires multiple model replicas, it first prioritizes using replicas within the same node. If those are insufficient, cross-node replicas are employed. When multiple combinations of replicas can satisfy the requirement, the combination with the smallest total local queue length is selected.

\noindent\textbf{Short request decode.}
Based on the resource consumption characteristics of short request prefill and decode~\cite{distserve,splitwise}, we allocated 4, 4, 1, and 1 dedicated model replicas for short request decode when evaluating \sys on Mistral-v0.3 7B, Phi-3 14B, Yi 34B, and Llama-3.1 70B, respectively.

\subsection{Overall Performance}

We measured the queueing delay and throughput of all short requests, as well as the JCT of long requests under each method. Fig.~\ref{fig:q_dealy} presents the 1st, 25th, 50th, 75th, and 99th percentile normalized queueing delays of all short requests. Across all models, the 99th percentile queueing delays of short requests under \sys and Priority are similar. This is because, under Priority, short requests are prioritized and thus not delayed by long requests. In \sys, short requests can preempt long request prefill, thereby avoiding delay. Compared to FIFO and Reservation, \sys reduces the 99th percentile queueing delay of short requests by 58\%–87\% and 61\%–92\%, respectively, across all models. This improvement arises because long requests delay short ones under FIFO, and in Reservation, fewer GPUs are reserved for short requests than for long ones, causing a huge number of short requests to experience delayed service. The larger the size of the model, the greater the reduction in queueing delay achieved by \sys, as longer execution times of long requests lead to more severe delays for short requests.

\noindent
\begin{minipage}[t]{0.32\textwidth}\vspace{-0.25in}
  \begin{figure}[H]
    \centering
    \includegraphics[width=\linewidth,height=2.3cm]{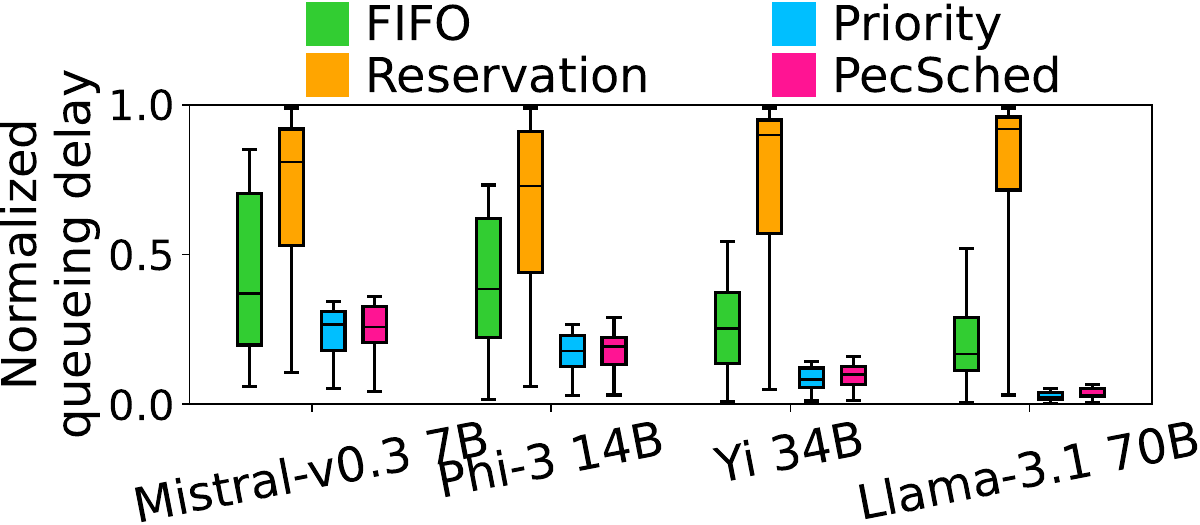}
    \vspace{-0.3in}
    \caption{Normalized queueing delay of short requests.}
    \label{fig:q_dealy}
    \vspace{-0in}
  \end{figure}
\end{minipage}%
\hspace{0.01\textwidth}
\begin{minipage}[t]{0.32\textwidth}\vspace{-0.25in}
  \begin{figure}[H]
    \centering
    \includegraphics[width=\linewidth,height=2.3cm]{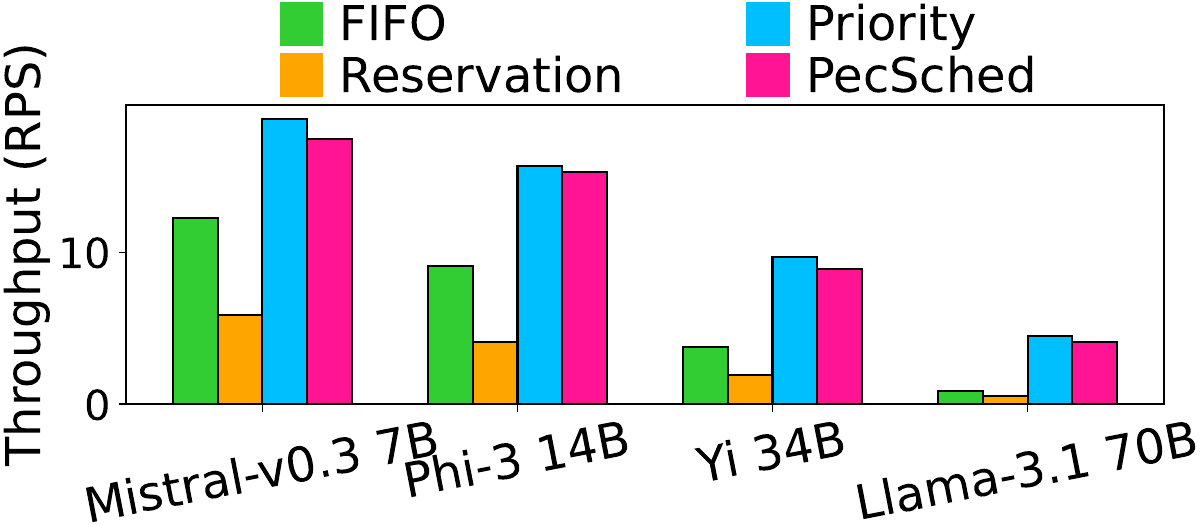}
    \vspace{-0.3in}
    \caption{Throughput of short requests.}
    \label{fig:throughput}
    \vspace{-0in}
  \end{figure}
\end{minipage}%
\hspace{0.01\textwidth}
\begin{minipage}[t]{0.32\textwidth}\vspace{-0.25in}
  \begin{figure}[H]
    \centering
    \includegraphics[width=\linewidth,height=2.3cm]{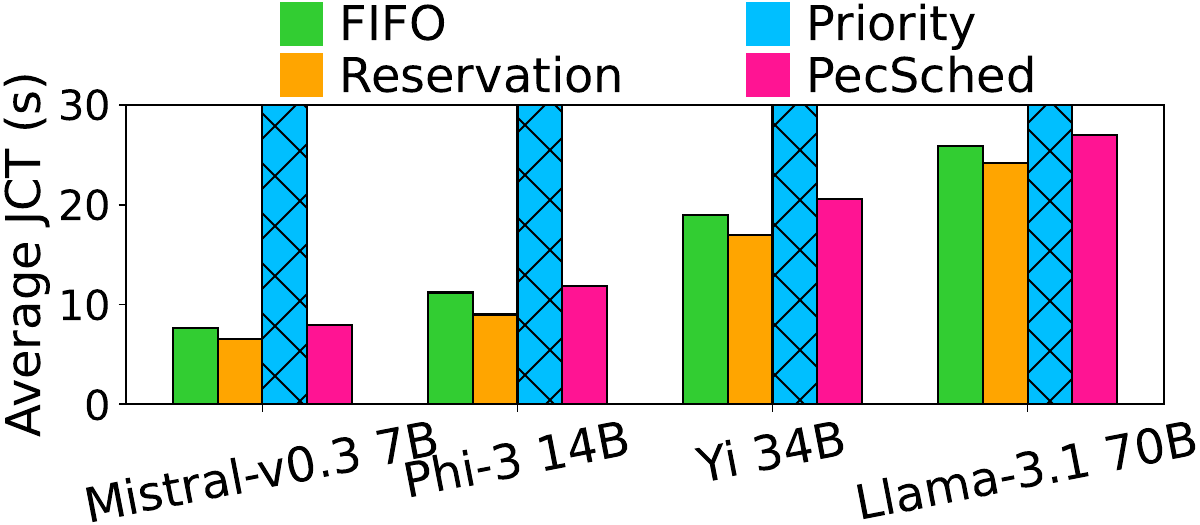}
    \vspace{-0.3in}
    \caption{Average JCT of long requests.}
    \label{fig:jct}
    \vspace{-0in}
  \end{figure}
\end{minipage}

Fig.~\ref{fig:throughput} shows the throughput of all short requests. For all models, the throughput under \sys is similar to that under Priority. Compared to FIFO and Reservation, \sys improves short request throughput by 42\%–318\% and 193\%–595\%, respectively. The gains grow with model size, due to the same reasons explained for queueing delay.

Fig.~\ref{fig:jct} shows the average JCT of all long requests. The average JCT under Priority is unbounded, as over 90\% of long requests are starved. In contrast, \sys increases the average JCT of long requests by only 4\%–7\% and 6\%–13\% compared to FIFO and Reservation, respectively. This is because \sys disaggregates the prefill and decode phases of short requests. Consequently, the queueing delay of long requests is only affected by the prefill phase of short requests, and the prefill of long requests can only be preempted by the prefill of short requests, not their decode. Moreover, colocating long request decode with short request prefill avoids interruptions to the former. The use of fast SP for long request prefill further reduces prefill time, thereby lowering both the probability and duration of preemptions. Since long requests have long JCTs, the slight increase in JCTs has a negligible impact on their overall performance.

\vspace{-0in}
\subsection{Ablation Study}

We test the variants of \sys as follows to evaluate each individual method. 1) \sys/PE is \sys without \underline{P}re\underline{E}mption. Short request prefill must wait for the completion of long request prefill before execution. 2) \sys/Dis is \sys without \underline{Dis}aggregation of short request decode. 3) \sys/CoL is \sys without \underline{CoL}ocation of long request decode and short request prefill. Short request prefill also preempts long request decode. 4) \sys/FSP is \sys without \underline{F}ast \underline{SP} for long request prefill.

\noindent
\begin{minipage}[t]{0.32\textwidth}\vspace{-0.25in}
  \begin{figure}[H]
    \centering
    \includegraphics[width=\linewidth,height=2.3cm]{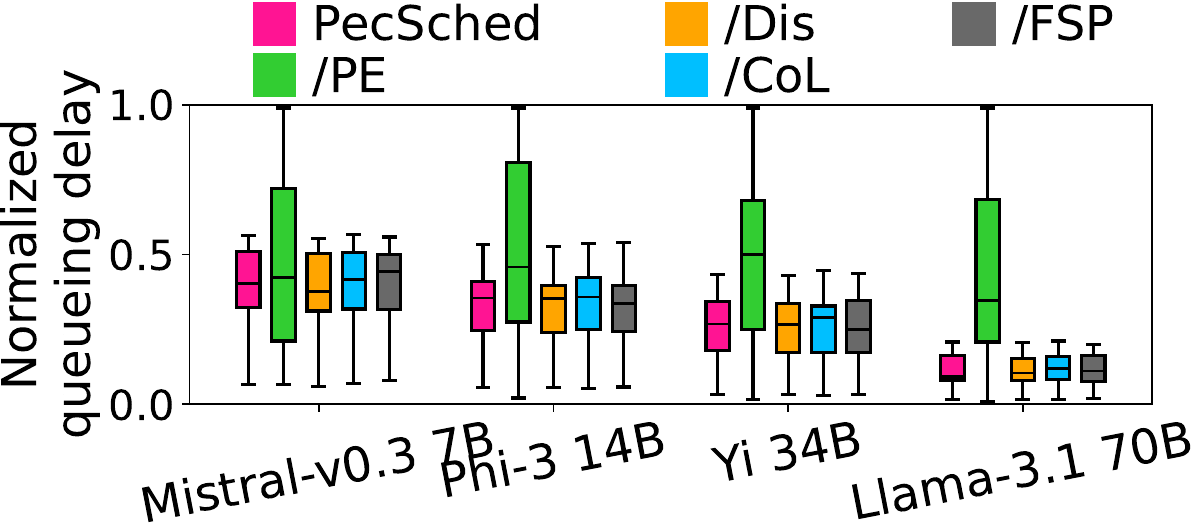}
    \vspace{-0.3in}
    \caption{Normalized queueing delay of short requests for individual methods.}
    \label{fig:q_dealy_indiv}
    \vspace{-0in}
  \end{figure}
\end{minipage}%
\hspace{0.01\textwidth}
\begin{minipage}[t]{0.32\textwidth}\vspace{-0.25in}
  \begin{figure}[H]
    \centering
    \includegraphics[width=\linewidth,height=2.3cm]{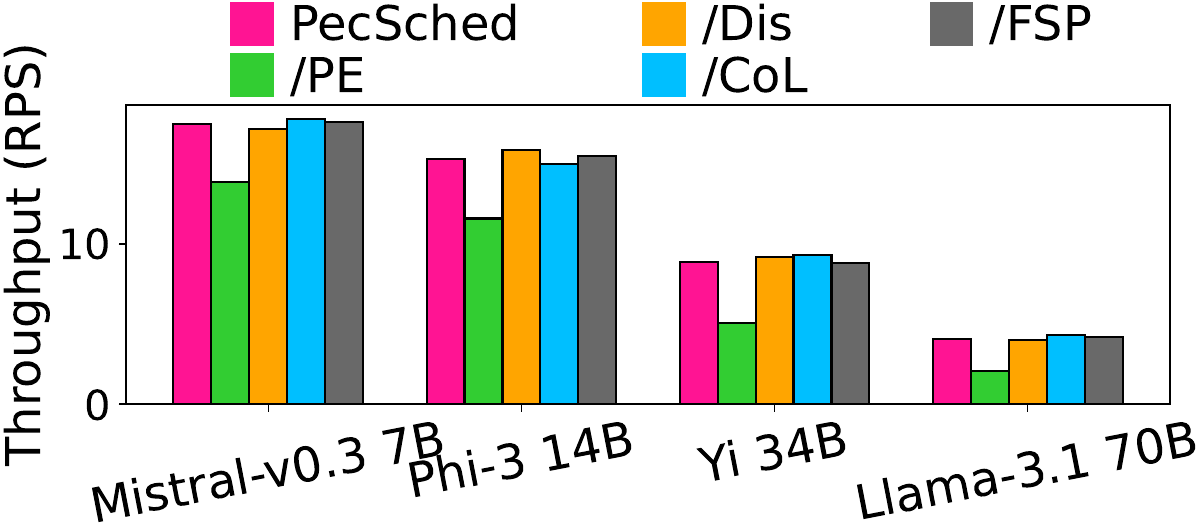}
    \vspace{-0.3in}
    \caption{Throughput of short requests for individual methods.}
    \label{fig:throughput_indiv}
    \vspace{-0in}
  \end{figure}
\end{minipage}%
\hspace{0.01\textwidth}
\begin{minipage}[t]{0.32\textwidth}\vspace{-0.25in}
  \begin{figure}[H]
    \centering
    \includegraphics[width=\linewidth,height=2.3cm]{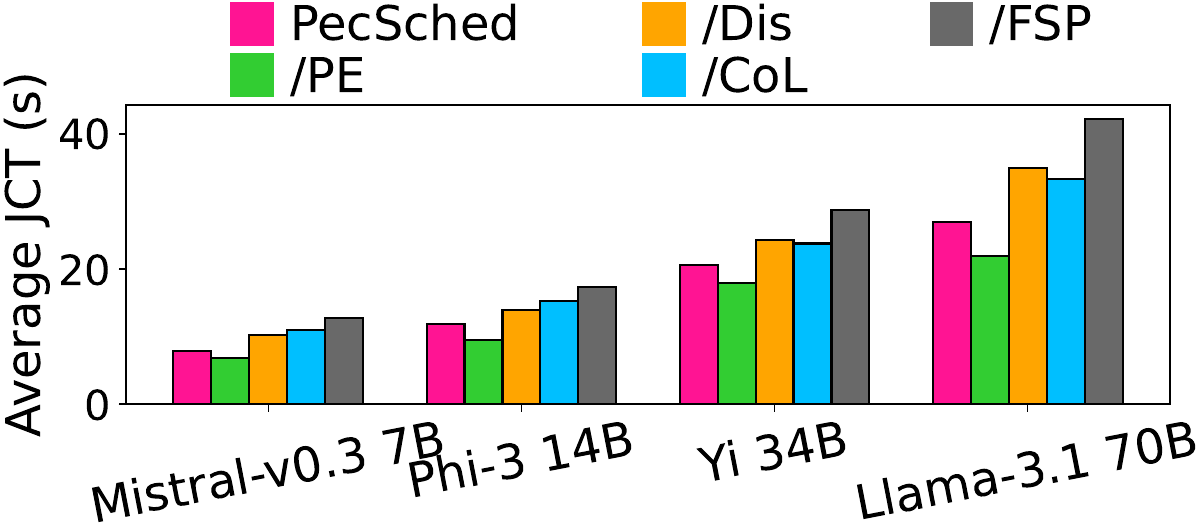}
    \vspace{-0.3in}
    \caption{Average JCT of long requests for individual methods.}
    \label{fig:jct_indiv}
    \vspace{-0in}
  \end{figure}
\end{minipage}

Fig.~\ref{fig:q_dealy_indiv} shows the 1st, 25th, 50th, 75th, and 99th percentile normalized queueing delays of short requests under each individual method.
The /PE method exhibits a 75\%–376\% higher 99th percentile queueing delay compared to \sys, as it lacks preemption. Short requests must wait for long request prefill to complete before execution, thereby increasing their queueing delay.
In contrast, other individual methods have similar 99th percentile queueing delays to \sys, as they still incorporate preemption mechanisms that prevent long requests from blocking the execution of short ones.
Fig.~\ref{fig:throughput_indiv} shows the throughput of short requests under different individual methods.
The /PE method achieves 21\%–48\% lower throughput compared to \sys. Other individual methods exhibit throughput similar to \sys. This is due to the same reasons discussed in the context of Fig.~\ref{fig:q_dealy_indiv}.
Fig.~\ref{fig:jct_indiv} presents the average JCT of long requests under different individual methods. The /PE method achieves 14\%–18\% lower average JCT compared to \sys, as short request prefill does not preempt long request prefill, avoiding suspension and thereby reducing the JCT of long requests. In contrast, /Dis, /CoL, and /FSP exhibit 21\%–29\%, 23\%–26\%, and 39\%–55\% higher average JCTs than \sys, respectively. This is because /Dis does not separate short request decode and prefill, allowing short request decode to preempt long request prefill and thus increasing its JCT. Under /CoL, short request prefill can preempt long request decode, also prolonging JCT. /FSP, on the other hand, increases the prefill time of long requests, which leads to a higher number and duration of preemptions by short requests.

Table~\ref{tab:num_preempt_indiv} shows the total number of preemptions experienced by long requests under the individual methods that have the preemption mechanism. As shown, /Dis, /CoL, and /FSP all incur more preemptions than \sys. For /Dis and /FSP, this is because the extended prefill phase increases the likelihood of being preempted. For /CoL, the increase is attributed to the fact that long request decode is also subject to preemption.

\begin{table}[h]\vspace{-0.1in}
\centering
\begin{tabular}{ |c|c|c|c|c|  }
 \hline
 & Mistral-v0.3 7B & Phi-3 14B & Yi 34B & Llama-3.1 70B\\
 \hline
 \sys & 94,057 & 116,290 & 139,247 & 170,914 \\
 \hline
 /Dis & 108,552 & 133,784 & 156,891 & 203,628 \\
 \hline
 /CoL & 130,925 & 167,607 & 209,834 & 261,720 \\
 \hline
 /FSP & 167,394 & 205,947 & 278,504 & 379,305 \\
 \hline
 \end{tabular}
\caption{The total number of preemptions experienced by all long requests for individual methods.}\label{tab:num_preempt_indiv}
\vspace{-0.3in}
\end{table}

% \begin{figure}[h]
%     \centering
%     \begin{minipage}[b]{0.31\columnwidth}
%         \includegraphics[width=\linewidth]{fig_conext25/q_delay.pdf}
%     \end{minipage}
%     \hspace{0.01\textwidth}
%     \begin{minipage}[b]{0.31\columnwidth}
%         \includegraphics[width=\linewidth]{fig_conext25/throughput.pdf}
%     \end{minipage}
%     \hspace{0.01\textwidth}
%     \begin{minipage}[b]{0.31\columnwidth}
%         \includegraphics[width=\linewidth]{fig_conext25/jct.pdf}
%     \end{minipage}
%     \caption{Three independent figures aligned in one row.}
% \end{figure}

\subsection{Time Overhead}

We evaluate the scheduling overheads for both short and long requests under \sys. For short requests, the scheduling time includes both the scheduling decision time and the context-switching time incurred when preempting the prefill of long requests. For long requests, it consists of the scheduling decision time and the time spent selecting the SP strategy for fast SP. Table~\ref{tab:overhead} reports the 99th percentile of the scheduling time ratio to JCT for short and long requests.
We observe that the ratio of scheduling time to JCT does not exceed 0.345\%, indicating that the scheduling overhead is acceptable. As the model size or request length increases, this ratio decreases because the JCT increases while the scheduling overhead remains relatively constant, resulting in a lower overhead-to-JCT ratio.

\begin{table}[h]\vspace{-0.1in}
\centering
\begin{tabular}{ |c|c|c|c|c|  }
 \hline
 & Mistral-v0.3 7B & Phi-3 14B & Yi 34B & Llama-3.1 70B\\
 \hline
 Short requests & 0.354\% & 0.282\% & 0.196\% & 0.071\% \\
 \hline
 Long requests & 0.183\% & 0.147\% & 0.055\% & 0.019\% \\
 \hline
 \end{tabular}
\caption{The 99th percentile ratio of scheduling time to JCT for long and short requests.}\label{tab:overhead}
\vspace{-0.3in}
\end{table}

\subsection{Scalability Test}

\begin{figure}[h] \vspace{-0in}
    \centering
    \includegraphics[width=0.66\columnwidth]{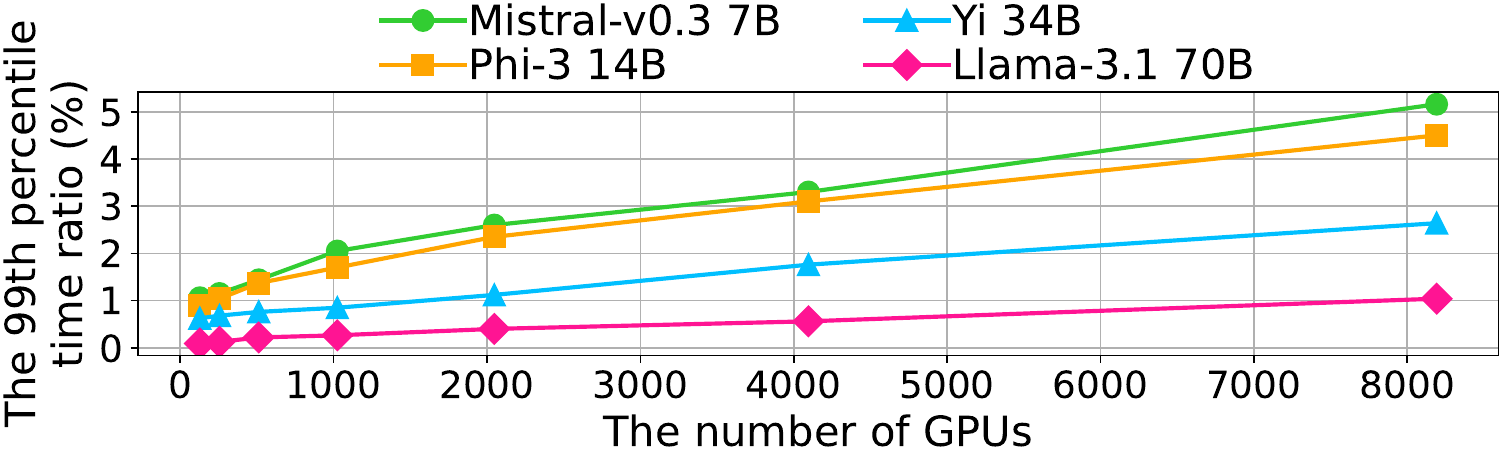}
    \vspace{-0.15in}
    \caption{The 99th percentile ratio of scheduling time to JCT with different numbers of GPUs.}
    \vspace{-0.1in}
    \label{fig:scale}
\end{figure}

We conduct a simulation-based study to evaluate the impact of cluster scale on the scheduling overhead of \sys. Based on the input and output lengths of requests in the trace we used in~\cref{sec:eval:setup}, we simulate their prefill and decode execution times. During the simulation, we vary the total number of GPUs in the cluster, while keeping each server node configured with 8 GPUs, consistent with~\cref{sec:eval:setup}.
For each GPU count setting, we generate requests following a Poisson distribution, with the request arrival rate (RPS) set to the cluster's maximum capacity based on the throughput data in Fig.~\ref{fig:throughput}.
Fig.~\ref{fig:scale} shows the 99th percentile of the scheduling time ratio to JCT under different GPU counts. We observe that as the number of GPUs increases, the 99th percentile scheduling time ratio grows approximately linearly. Even when the cluster size reaches 8192 GPUs, the ratio remains below 5.2\%, which is acceptable.
The scheduling time ratio decreases as model size increases. This is because larger models lead to higher JCTs, while the number of model replicas that influence the scheduling search space becomes smaller, which in turn slightly reduces the scheduling decision time. For example, the scheduling time ratio for Llama-3.1 70B remains below 1.1\% in the simulation. Given that modern clusters are typically dominated by large models, we expect the scheduling overhead of \sys to remain low in practice.

\vspace{0in}

% \begin{table}[h]
% \centering
% \begin{tabular}{ |c|c|c|c|c|c|c|c|c|  }
%  \hline
%   & \# of GPUs & 128 & 256 & 512 & 1024 & 2048 & 4096 & 8192 \\
%  \hline
%  \multirow{2}{*}{Mistral-v0.3 7B} & RPS & 128 & 256 & 512 & 1024 & 2048 & 4096 & 8192 \\
%  \cline{2-9}
%   & Ratio & 0.354\% & 0.282\% & 0.196\% & 0.071\% & 0.196\% & 0.071\% & 0.071\% \\
%  \hline
%  \multirow{2}{*}{Phi-3 14B} & RPS & 128 & 256 & 512 & 1024 & 2048 & 4096 & 8192 \\
%  \cline{2-9}
%   & Ratio & 0.354\% & 0.282\% & 0.196\% & 0.071\% & 0.196\% & 0.071\% & 0.071\% \\
%  \hline
%  \multirow{2}{*}{Yi 34B} & RPS & 128 & 256 & 512 & 1024 & 2048 & 4096 & 8192 \\
%  \cline{2-9}
%   & Ratio & 0.354\% & 0.282\% & 0.196\% & 0.071\% & 0.196\% & 0.071\% & 0.071\% \\
%  \hline
%  \multirow{2}{*}{Llama-3.1 70B} & RPS & 128 & 256 & 512 & 1024 & 2048 & 4096 & 8192 \\
%  \cline{2-9}
%   & Ratio & 0.354\% & 0.282\% & 0.196\% & 0.071\% & 0.196\% & 0.071\% & 0.071\% \\
%  \hline
%  \end{tabular}
% \caption{The average ratio of scheduling time to JCT for long and short requests.}\label{tab:scale}
% \end{table}

\vspace{-0.0in}
\section{Related Work}\label{related-work}\vspace{-0in}

\noindent\textbf{Cluster-level LLM inference scheduler.}
Current cluster-level LLM inference scheduling strategies can be broadly categorized into three classes: FIFO-based~\cite{vllm2023kwon, loongserve}, reservation-based~\cite{llumnix}, and priority-based~\cite{exegpt,past-future,fastserve} approaches. FIFO-based strategies serve requests in the order of their arrival. Reservation-based strategies allocate dedicated resources to different types of requests to avoid resource interference. Priority-based strategies assign higher priorities to short-output requests to improve their throughput. However, none of these approaches account for the coexistence of short-input and long-input requests within a cluster, which can lead to head-of-line blocking, low resource utilization, and starvation of long-input requests. Our system addresses these limitations.

\noindent\textbf{Inference execution enhancement.}
Recent efforts were made to enhance LLM inference execution.
vLLM~\cite{vllm2023kwon} uses Paged Attention to enable a non-contiguous KV cache, which reduces memory fragmentation and increases throughput. HuggingFace TGI~\cite{hf-tgi} and NVIDIA TensorRT-LLM~\cite{tensorrt-llm} have also implemented the non-contiguous KV cache.
DistServe~\cite{distserve} and SplitWise~\cite{splitwise} split prefill and decode to improve throughput.
FastGen~\cite{holmes2024deepspeedfastgen} and Sarathi-Serve~\cite{Agrawal2023SARATHIEL} chunk a long prompt and batch chunks sequentially with token generation tasks.
These works focus on improving inference execution rather than request scheduling within the cluster.

\noindent\textbf{Sequence parallelism for long request execution.}
SP~\cite{korthikanti2022reducing, deepspeed-ulysses, li2021sequence, ringattention} was initially proposed to train long-context models. Other recent studies~\cite{wang2020linformer, winata2020lightweight, katharopoulos2020transformers, choromanski2020rethinking, qin2022devil, chaudhuri2019set-transformer, jaegle2021perceiver, ma2021luna, dai2019transformer-xl, bulatov2023scaling, wu2022memorizing, wang2023augmenting, ding2023longnet} proposed transformer variants to handle long sequences for training, striking a balance between performance and accuracy.
Recent work~\cite{loongserve} introduced SP with ring attention~\cite{ringattention} to LLM inference to handle long requests losslessly. However, it does not address the scheduling of short and long requests in a cluster. Our system handles this.\looseness=-1

\DEL{\noindent \textbf{Token importance.}
Recent studies~\cite{guo2021longt5, ding2023longnet} have shown that nearby tokens are important to each other and distant tokens are less important to each other for long prompts. Other studies~\cite{h2o2023zhang,scissorhands2023liu,ge2023model} investigated and identified token importance using attention scores. They find unimportant tokens and drop those tokens to save KV cache for inference tasks.}

\vspace{-0.0in}
\section{Discussion and Limitations}\label{discussion}\vspace{-0in}

\noindent\textbf{Colocation of long request prefill with short request decode.}
We do not consider colocating long request prefill with short request decode because the long prefill time of long requests can delay a decode iteration of short requests. Additionally, the total decode time for short requests varies across different model replicas and is unpredictable, further complicating colocation.

\noindent\textbf{Limitations and future work.}
% Although \sys reduces the impact of preemption on long requests, it does not completely eliminate the associated overhead. The prefill phase of long requests may still be suspended, introducing some degree of delay. In the future, we will explore better approaches to enhance the throughput of short requests while further reducing the JCT of long requests.
Although \sys mitigates preemption overhead for long requests, it does not fully eliminate it, as long request prefill phases may still be suspended. Our future work will explore improved methods to boost short request throughput while further reducing long request JCT.

\vspace{-0.0in}
\section{Conclusion}\label{conclusion}\vspace{-0in}

This paper presents \sys, a preemptive and efficient cluster scheduling system for LLM inference. \sys introduces preemptive scheduling, prefill-decode disaggregation of short requests, colocation of long request decode with short request prefill, and fast SP with a hybrid SP strategy for long request prefill. It improves the performance of short requests without significantly affecting the JCT of long requests. Both real experiments and large-scale simulations show the superior performance of \sys in comparison with the state-of-the-art. \looseness=-1

%applying \methodone that considers the SPT placement on GPUs and efficiently identifies and drops unimportant tokens from far away tokens a token attends to. We designed and implemented an \methodone-based pipeline to overlap the SP communication and computation. We also built two regression models for the objective-oriented hyper-parameter configuring. The evaluation shows that \Sys improves the average TBT by up to 1.92$\times$, reduces average response time by up to 9.8$\times$, and improves the request rate by up to 2.5$\times$ while keeping a similar latency under the inference accuracy degradation constraint. \Sys can also effectively minimize the accuracy loss under the TTFT constraint.

\newpage

\bibliographystyle{unsrt}
\bibliography{reference}

\end{document}